\documentclass{ws-procs9x6}

\begin{document}
\newcommand{\be}{\begin{equation}}
\newcommand{\ee}{\end{equation}}
\newcommand{\bea}{\begin{eqnarray}}
\newcommand{\eea}{\end{eqnarray}} \newcommand{\nn}{\nonumber}
\newcommand{\dd}{\displaystyle}
\newcommand{\de}{\partial}
\title{On the Ground State of QCD inside a Compact Stellar Object}

\author{Roberto Casalbuoni}

\address{Department of Physics, University of Florence,\\
Florence, Italy and\\
Sezione INFN, Florence, Italy\\
E-mail: casalbuoni@fi.infn.it\\
www.theory.fi.infn.it/casalbuoni/ }

\begin{abstract}
We describe the effects of the strange quark mass and of the color
and electric neutrality on the superconducing phases of QCD.
\end{abstract}


\bodymatter

\section{Introduction}

It is now a well established fact that at zero temperature and
sufficiently high densities quark matter is a color superconductor
\cite{barrois,cs}\,. The study starting from first principles was
done in Refs. \cite{weak,PR-sp1,weak-cfl}\,. At baryon chemical
potentials much higher than the masses of the quarks $u$, $d$ and
$s$, the favored state is the so-called Color-Flavor-Locking (CFL)
state, whereas at lower values, when the strange quark decouples,
the relevant phase is called two-flavor color superconducting (2SC).

An interesting possibility is that in the interior of compact
stellar objects (CSO) some color superconducting phase may exist. In
fact the central densities for these stars could be up to $10^{15}$
g/cm$^{3}$, whereas the temperature is of the order of tens of keV.
However the usual assumptions leading to prove that for three
flavors the favored state is CFL should now be reviewed. Matter
inside a CSO should be electrically neutral and should not carry
color. Also  conditions for $\beta$-equilibrium should be fulfilled.
As far as color is concerned, it is possible to impose a simpler
condition, that is color neutrality, since in Ref.
\cite{Amore:2001uf} it has been shown that there is no free energy
cost in projecting color singlet states out of color neutral ones.
Furthermore one has to take into account that at the interesting
density the mass of the strange quark is a relevant parameter. All
these effects, the mass of the strange quark, $\beta$-equilibrium
and color and electric neutrality, imply that the radii of the Fermi
spheres of  quarks that would pair are not the same. This difference
in radius, as we shall see, is going to create a problem with the
usual BCS pairing. Let us start from the mass effects. Suppose to
have two fermions of masses $m_1=M$ and $m_2=0$ at the same chemical
potential $\mu$. The corresponding Fermi momenta are
$p_{F_1}=\sqrt{\mu^2-M^2}$ and $p_{F_2}=\mu$. We see that the radius
of the Fermi sphere of the massive fermion is smaller than the one
of the massless particle. If we assume $M\ll \mu$  the massive
particle has an effective chemical potential
$\mu_{\rm{eff}}=\sqrt{\mu^2-M^2}\approx \mu-{M^2}/{2\mu}$ and the
mismatch between the two Fermi spheres is given by
\be\delta\mu\approx\frac{M^2}{2\mu}\label{eq:3}\ee This shows that
the quantity $M^2/(2\mu)$ behaves as a chemical potential. Therefore
for $M\ll\mu$  the mass effects can be taken into account through
the introduction of the  mismatch between the chemical potentials of
the two fermions given by eq. (\ref{eq:3}). This is  the way that we
will follow in our study.

Now let us discuss  $\beta$-equilibrium. If electrons are present
(as generally required by electrical neutrality) chemical potentials
of quarks of different electric charge are different. In fact, when
at the equilibrium for $d\to ue\bar\nu$, we have \be
\mu_d-\mu_u=\mu_e\ee From this condition it follows that for a quark
of charge $Q_i$ the chemical potential $\mu_i$ is given by \be
\mu_i=\mu+Q_i\mu_Q\ee where $\mu_Q$ is the chemical potential
associated to the electric charge. Therefore \be\mu_e=-\mu_Q\ee
Notice also that $\mu_e$ is not a free parameter since it is
determined by the neutrality condition \be Q=-\frac{\de \Omega}{\de
\mu_e}=0\ee At the same time the chemical potentials associated to
the color generators $T_3$ and $T_8$ are determined by the color
neutrality conditions \be\frac{\de \Omega}{\de \mu_3}= \frac{\de
\Omega}{\de \mu_8}=0\label{eq:1}\ee

We see that in general there is a mismatch between the quarks that
should pair according to the BCS mechanism at $\delta\mu=0$.
Increasing the mismatch has the effect of destroying the BCS phase
and the system either goes into the normal phase or to some
different phase. In the next Sections we will explore some of these
possible alternatives.

\section{Neutrality and $\beta$-equilibrium}

Just as a very simple example of the effect of the neutrality and
$\beta$-equilibrium conditions, let us consider three non
interacting quarks, $u$, $d$ and $s$. The $\beta$-equilibrium
requires \be \mu_{d,s}=\mu_u+\mu_e\ee The chemical potentials of the
single species in term of the baryon chemical potential, $\bar\mu$,
and of the charge chemical potential, $\mu_Q=-\mu_e$, are therefore
\be \mu_u=\bar\mu-\frac 23\mu_e,~~~\mu_d=\mu_s=\bar\mu+\frac 1 3
\mu_e\ee The numerical densities of different quarks are given by
\be N_{u,d}=\frac{\mu_{u,d}^3}{\pi^2},~~~
N_s=\frac{(\mu_s^2-M_s^2)^{3/2}}{\pi^2},~~~N_e=\frac{\mu_e^3}{3\pi^2}\ee
\begin{figure}[b]\begin{center}
  \includegraphics[height=.3\textheight]{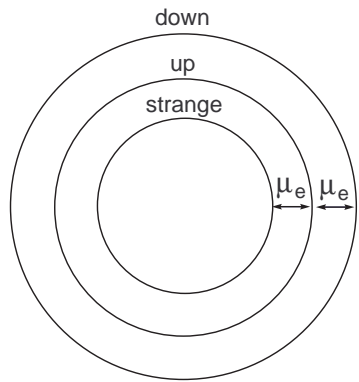}
  \caption{The Fermi spheres for three non interacting quarks, $u$,
  $d$ and $s$ by taking into account the mass of the strange quark
  (see text).
  \label{fig:1}}\end{center}
\end{figure}
On the other hand the neutrality condition requires \be \frac 2 3
N_u-\frac 1 3 N_d-\frac 13 N_s-N_e=0\ee If the strange quark mass is
neglected the previous equation has the simple solution \be
N_u=N_d=N_s,~~~N_e=0\ee In this case the Fermi spheres of the three
quarks have the same radius (remember that for a single fermion  the
numerical density is given by $N=p_F^3/(3\pi^2)$). However if we
take into account $M_s\not= 0$ at the lowest order in $M_s/\mu$ we
get \be \mu_e\approx\frac{M_s^2}{4\mu},~~~ p_F^d-p_F^u\approx
p_F^u-p_F^s\approx\mu_e\ee The result is shown in Fig. \ref{fig:1}.
It can also be shown that in the normal phase the chemical
potentials associated to the color charges $T_3$ and $T_8$ vanish.
We will make use of these results when we will discuss the LOFF
phase.

\section{Gapless quasi-fermions}

When a  mismatch is present, the spectrum of the quasi-particles is
modified as follows\be E_{\delta\mu=0}=\sqrt{(p-\mu)^2+\Delta^2}\to
E_{\delta\mu}=\left|\delta\mu\pm\sqrt{(p-\mu)^2+\Delta^2}\,\right|\ee
Therefore for $|\delta\mu|<\Delta$ we have gapped quasi-particles
with gaps $\Delta\pm\delta\mu$. However, for $|\delta\mu|=\Delta$ a
gapless mode appears and from this point on there are regions of the
phase space which do not contribute to the gap equation (blocking
region).

The gapless modes are characterized by \be E(p)=0\Rightarrow
p=\mu\pm\sqrt{\delta\mu^2-\Delta^2}\ee Since the energy cost for
pairing two fermions belonging to Fermi spheres with mismatch
$\delta\mu$ is $2\delta\mu$ and the energy gained in pairing is
$2\Delta$, we see that the fermions begin to unpair for $
2\delta\mu\ge 2\Delta$. These considerations will be relevant for
the study of the gapless phases when neutrality is required.

\section{The gCFL phase}

The gCFL phase is a generalization of the CFL phase which has been
studied both at $T=0$ \cite{Alford:2003fq,Alford:2004hz} and
$T\not=0$ \cite{gCFL_2}. The condensate has now the form \be \langle
0|\psi_{aL}^\alpha\psi_{bL}^\beta|0\rangle=\Delta_1\epsilon^{\alpha\beta
1}\epsilon_{ab1}+\Delta_2\epsilon^{\alpha\beta
2}\epsilon_{ab2}+\Delta_3\epsilon^{\alpha\beta 3}\epsilon_{ab3}\ee
The CFL phase corresponds to all the three gaps $\Delta_i$ being
equal. Varying the gaps one gets many different phases. In
particular we will be interested to the CFL, to the g2SC
characterized by $\Delta_3\not=0$ and $\Delta_1=\Delta_2=0$ and to
the gCFL phase with $\Delta_3>\Delta_2>\Delta_1$. Notice that in the
g2SC phase defined here the strange quark is present but unpaired.

In flavor space the gaps $\Delta_i$ correspond to the following
pairings in flavor\be \Delta_1\Rightarrow ds,~~~\Delta_2\Rightarrow
us,~~~ \Delta_3\Rightarrow ud\ee The mass of the strange quark is
taken into account by shifting all the chemical potentials involving
the strange quark as follows: $\mu_{\alpha s}\to \mu_{\alpha s}
-{M_s^2}/{2\mu}$. It has also been shown in ref. \cite{alford} that
color and electric neutrality in CFL require \be
\mu_8=-\frac{M_s^2}{2\mu},~~~\mu_e=\mu_3=0\ee At the same time the
various mismatches are given by \be
\delta\mu_{bd-gs}=\frac{M_s^2}{2\mu},~~~\delta\mu_{rd-gu}=\mu_e=0,~~~
\delta\mu_{rs-bu}=\mu_e-\frac{M_s^2}{2\mu}\ee It turns out that in
the gCFL the electron density is different from zero and, as a
consequence, the mismatch between the quarks $d$ and $s$ is the
first one to give rise to the unpairing of the corresponding quarks.
This unpairing is expected to occur for \be
2\frac{M_s^2}{2\mu}>2\Delta~~
\Rightarrow~~\frac{M_s^2}{\mu}>2\Delta\ee This has been
substantiated in \cite{Alford:2004hz} by a calculation in the NJL
model based on one gluon-exchange. The authors assume for their
calculation a chemical potential, $\mu=500~MeV$ and a CFL gap given
by $\Delta=25~MeV$. The transition from the CFL phase, where all
gaps are equal, to the gapless phase occurs roughly at $M_s^2/\mu
=2\Delta$.
\begin{figure}[h]\begin{center}
  \includegraphics[height=.3\textheight]{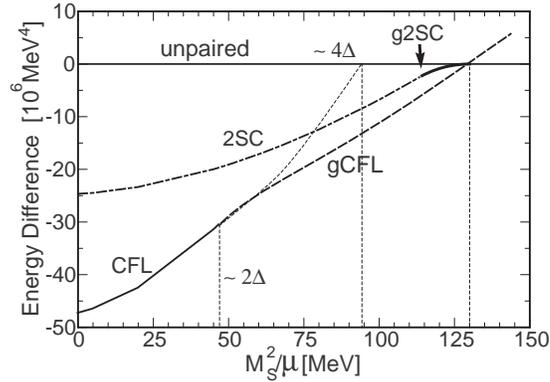}
  \caption{Free energy of the various phases discussed in the text with
  reference to the normal phase, named unpaired in the figure.
  \label{fig:2}}\end{center}
\end{figure}
In Fig. \ref{fig:2} we show the free energy of the various phases
with reference to the normal phase. The CFL phase is the stable one
up to $M_s^2/\mu\approx  2\Delta$. Then the gCFL phase takes over up
to about 130 $MeV$, where the system goes to the normal phase.
Notice that except in a very tiny region around this point, the CFL
and gCFL phases dominate over the corresponding 2SC and g2SC ones.
The thin short-dashed line represents the free energy of the CFL
phase up to the point where it becomes equal to the free-energy of
the normal phase. This happens for $M_s^2/\mu\approx 4\Delta$.

Although the gCFL phase appears to be energetically favored it
cannot be the real ground state. In fact, it has been shown in
\cite{MeissnerCFL_1,MeissnerCFL_2} that in this phase there is a
chromomagnetic instability. This instability manifests itself in the
masses of the gluons 1, 2, 3, 8 becoming pure imaginary at the
transition CFL-gCFL. An analogous  instability (relative to the
gluons 4, 5, 6, 7, 8) occurs in the g2SC phase
 \cite{huang_instability1,Hashimoto:2006mn,Kiriyama:2006xw,Kiriyama:2006jp}
  and it seems to be related to the gapless modes present in
 homogeneous phases, as conjectured in Ref.
\cite{alford_wang}\,.

\section{Possible solutions of the problem of the chromagnetic instability}

There have been various proposals to solve the problem of the
chromo-magnetic instability. We will shortly review these attempts
before discussing the proposal that at the moment seems to be the
favored one, that is the one corresponding to the LOFF phase (see
next Section):
\begin{itemize}
\item {\bf Gluon condensation}. If one assumes artificially that the
expectation values of $A_{\mu 3}$ and $A_{\mu 8}$ are not zero, and
of the order of 10 $MeV$, the instability goes away
\cite{MeissnerCFL_1}\,. This argument has been done more accurate
for the g2SC phase in Refs.
\cite{Gorbar:2005rx,Gorbar:2005tx,Kiriyama:2006ui,He:2006vr}\,,
where it has been considered a model exhibiting  chromo-magnetic
condensation. It turns out that the rotational symmetry is broken
and this makes some connection with the LOFF phase. At the moment
these models have not been extended to the three flavor case.
\item{\bf CFL-K$^0$ phase}. If the mismatch is not too large
(meaning $\delta\mu/\mu\ll 1$) the CFL pattern can be modified by a
flavor rotation of the condensate. This is equivalent to have a
condensate of kaons \cite{Bedaque:2001je}\,. The transition to this
phase occurs roughly for a strange quark mass satisfying
$M_s>m^{1/3}\Delta^{2/3}$, with $m$ the light quark mass and
$\Delta$ the CFL condensate. Also this phase exhibits gapless modes
and  the gluon instability occurs
\cite{Kryjevski:2004jw,Kryjevski:2004kt,Kryjevski:2005qq}\,.
Allowing for a space dependent condensate, a current is generated
which eliminates the instability \cite{Gerhold:2006dt}\,. Also in
this case, a space dependent condensation brings a relation to the
LOFF phase.
\item {\bf Single flavor pairing}. If the stress caused by the
mismatch is too big, single flavor pairing could occur. However the
gap appears to be too small. It could be important at low chemical
potential before the nuclear phase (see, for instance Ref.
\cite{Alford:2006wn}\,).
\item {\bf Secondary pairing}. The gapless modes could form a
secondary gap, but here too the gap is far too small
\cite{Shovkovy:2003ke,Alford:2005kj}\,.
\item {\bf Mixed phases}. The possibility of mixed phases both of
nuclear and quark matter \cite{Alford:2001zr} as well as mixed
phases of different Color Superconducting
\cite{Buballa:2003df,Alford:2004nf} phases has been considered.
However all these possibilities are either unstable or energetically
disfavored.
\item {LOFF phase}. In Ref. \cite{Ren1} it has been shown that the
chromagnetic instability of the g2SC phase is just what is needed in
order to make the crystalline, or LOFF phase, energetically favored.
Also it turns out that in the LOFF phase there is no chromomagnetic
instability although gapless modes are present \cite{Ren2}\,.
\end{itemize}
The previous considerations make the LOFF phase worth to be
considered and this is what we will do in the next Section.

\section{The LOFF Phase}

According to the authors of Refs. \cite{LOFF1,LOFF2} when fermions
belong to  different Fermi spheres, they  might prefer to pair
staying as much as possible close to their own Fermi surface. The
total momentum of the pair is not zero, ${\vec p}_1+{\vec p}_2=2\vec
q$ and, as we shall show, $|\vec q\,|$ is fixed variationally
whereas the direction of $\vec q$ is chosen spontaneously. Since the
total momentum of the pair is not zero the condensate breaks
rotational and translational invariance. The simplest form of the
condensate compatible with this breaking is just a simple plane wave
(more complicated possibilities will be discussed later) \be
\langle\psi(x)\psi(x)\rangle\approx\Delta\, e^{2i\vec q\cdot\vec
x}\label{single-wave}\ee It should also be noticed that the pairs
use much less of the Fermi surface than they do in the BCS case. For
instance, if both fermions are sitting at their own Fermi surface,
they can pair only if they belong to  circles fixed by $\vec q$.
More generally there is a quite large region in momentum space (the
so called blocking region) which is excluded from pairing. This
leads to a condensate generally smaller than the BCS one.

Let us now consider in more detail the LOFF phase (for reviews of
this phase see Refs.
\cite{Alford:2000ze,Bowers:2002xr,Casalbuoni:2003wh,Bowers:2003ye}).
For two fermions at different densities  we have an extra term in
the hamiltonian which can be written as \be
H_I=-\delta\mu\sigma_3\label{interaction}\ee where, in the original
LOFF papers \cite{LOFF1,LOFF2}\,, $\delta\mu$ is proportional to the
magnetic field due to the impurities, whereas in the actual case
$\delta\mu=(\mu_1-\mu_2)/2$ and $\sigma_3$ is a Pauli matrix acting
on the two fermion space. According to Refs. \cite{LOFF1,LOFF2} this
favors the formation of pairs with momenta \be \vec p_1=\vec k+\vec
q,~~~\vec p_2=-\vec k+\vec q\ee We will discuss in detail the case
of a single plane wave (see eq. (\ref{single-wave})). The
interaction term of eq. (\ref{interaction}) gives rise to a shift in
the quasi-particles energy  due both to the non-zero momentum of the
pair and to the different chemical potentials \be E(\vec p)-\mu\to
E(\pm\vec k+\vec q)-\mu\mp\delta\mu\approx E(\vec p)\mp\bar\mu\ee
with \be \bar\mu=\delta\mu-{\vec v}_F\cdot\vec q\ee Notice that the
previous dispersion relations show the presence of gapless modes at
momenta depending on the angle of $\vec v_F$ with $\vec q$. Here we
have assumed $\delta\mu\ll\mu$ (with $\mu=(\mu_1+\mu_2)/2$) allowing
us to expand $E$ at the first order in $\vec q/\mu$.

The study of the gap equation  shows that increasing $\delta\mu$
from zero we get first the BCS phase. Then at
$\delta\mu=\delta\mu_1$ there is a first order transition to the
LOFF phase \cite{LOFF1,Alford:2000ze}\,, and at
$\delta\mu=\delta\mu_2>\delta\mu_1$ there is a second order phase
transition to the normal phase \cite{LOFF1,Alford:2000ze}\,. We
start comparing the grand potential in the BCS phase to the one in
the normal phase. Their difference is given by (see for example Ref.
\cite{Casalbuoni:2003wh})\be \Omega_{\rm BCS}-\Omega_{\rm
normal}=-\frac{p_F^2}{4\pi^2v_F}\left(\Delta^2_0-2\delta\mu^2\right)\ee
where the first term comes from the energy necessary to the BCS
condensation, whereas the last term arises from the grand potential
of two free fermions with different chemical potential. We recall
also that for massless fermions $p_F=\mu$ and $v_F=1$. We have again
assumed $\delta\mu\ll\mu$. This implies that there should be a first
order phase transition from the BCS to the normal phase at
$\delta\mu=\Delta_0/\sqrt{2}$\, \cite{Chandrasekhar}\,, since the
BCS gap does not depend on $\delta\mu$. In order to compare with the
LOFF phase one can  expand the gap equation around the point
$\Delta=0$ (Ginzburg-Landau expansion)  to explore the possibility
of a second order phase transition \cite{LOFF1}\,. The result for
the free energy is \be \Omega_{\rm LOFF}-\Omega_{\rm normal}\approx
-0.44\,\rho(\delta\mu-\delta\mu_2)^2\ee At the same time, looking at
the minimum in $q$ of the free energy one finds \be qv_F\approx
1.2\, \delta\mu\label{q}\ee Since we are expanding in $\Delta$, in
order to get this result
 it is enough to minimize the
coefficient of $\Delta^2$ in the free-energy (the first term in the
Ginzburg-Landau expansion).

 We see that in the window
between the intersection of the BCS curve and the LOFF curve  and
$\delta\mu_2$, the LOFF phase is favored. Also at the intersection
there is a first order transition between the LOFF and the BCS
phase. Furthermore, since $\delta\mu_2$ is very close to
$\delta\mu_1$ the intersection point is practically given by
$\delta\mu_1$. The window of existence of the LOFF phase
$(\delta\mu_1,\delta\mu_2)\simeq(0.707,0.754)\Delta_0$ is rather
narrow, but there are indications that considering the realistic
case of QCD \cite{Leibovich:2001xr} the window  opens up. Such
opening occurs also for different crystalline structures than the
single plane wave\cite{Bowers:2002xr,Casalbuoni:2004wm}\,.

\section{The LOFF phase with three flavors}

\begin{figure}[h]\begin{center}
  \includegraphics[height=.2\textheight]{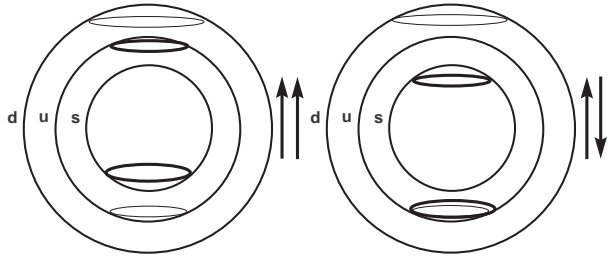}
  \caption{The two Fermi spheres corresponding to $q_2$ (left arrow) and $q_3$ (right
  arrow) respectively parallel and antiparallel. The pairing rings $du$ and $us$ are
  shown by thin and thick lines respectively.\label{fig:3}}\end{center}
\end{figure}

\begin{figure}[t]\begin{center}
\includegraphics[height=.3\textheight]{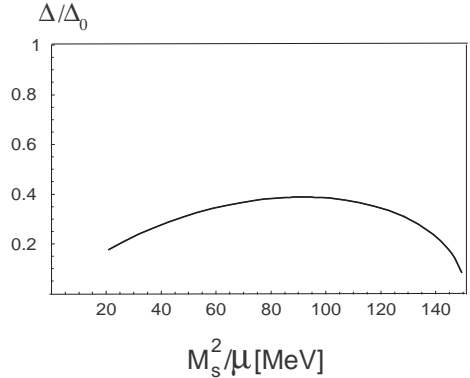}
\caption{The ratio of the gap $\Delta/\Delta_0$ for LOFF  with three
flavors vs. $M_s^2/\mu$. Here $\Delta_0$ is the CFL gap and
$\Delta=\Delta_2=\Delta_3$.
 \label{fig:4}}\end{center}
\end{figure}

\begin{figure}[h]\begin{center}
\includegraphics[height=.3\textheight]{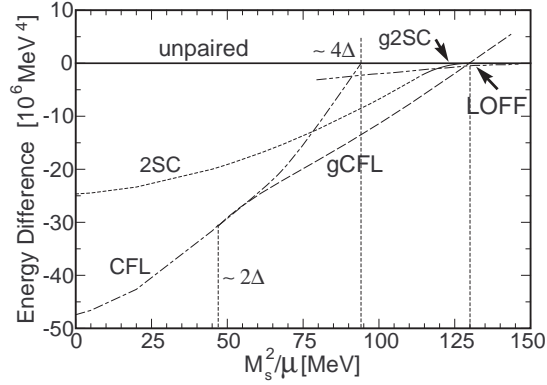}
\caption{Comparison of the free energy of the various phases already
considered in Fig. \ref{fig:2} (same notations as here) with the
LOFF phase with three flavors. \label{fig:5}}\end{center}
\end{figure}

In the last Section we would like to illustrate some preliminary
result about the LOFF phase with three flavors. This problem has
been considered in \cite{casalbuoni_loff3} under various simplifying
hypothesis:
\begin{itemize}
  \item The study has been made in the Ginzburg-Landau
  approximation.
  \item Only electrical neutrality has been required and the
  chemical potentials for the color charges $T_3$ and $T_8$ have
  been put equal to zero (see later).
  \item The mass of the strange quark has been introduced as it was
  done previously  for the gCFL phase.
  \item The study has been restricted to plane waves, assuming the
  following generalization of the gCFL case:
  \be \langle\psi^\alpha_{aL}\psi^\beta_{bL}\rangle=\sum_{I=1}^3\Delta_I(\vec x)
\epsilon^{\alpha\beta I}\epsilon_{ab I},~~~\Delta_I(\vec x)=\Delta_I
e^{2i\vec q_I\cdot\vec x}\label{simple_ansatz}\ee
\item The condensate depends on three momenta, meaning three lengths
of the momenta $q_i$ and three angles. In \cite{casalbuoni_loff3}
only  four particular geometries have been considered: 1) all the
momenta parallel, 2) $\vec q_1$ antiparallel to $\vec q_2$ and $\vec
q_3$, 3) $\vec q_2$ antiparallel to $\vec q_1$ and $\vec q_3$, 4)
$\vec q_3$ antiparallel to $\vec q_1$ and $\vec q_2$.
\end{itemize}
The minimization of the free energy with respect to the $|\vec
q_I|$'s leads to the same result as in eq. (\ref{q}), $ |\vec
q_I|=1.2 \delta\mu_I\label{eq:55}$. Let us notice that consistently
with the Ginzburg-landau approximation requiring to be close to the
normal phase, we assume $\mu_3=\mu_8=0$ as discussed in Section 2.
We remember also that close to the normal phase the Fermi surfaces
are given in Fig. \ref{fig:1} and as a consequence at the same order
of approximation we expect $\Delta_2=\Delta_3$ (since $ud$ and $us$
mismatches are equal) and $\Delta_1=0$, due to the $sd$ mismatch
being the double of the other two. Once we assume $\Delta_1=0$ only
the two configurations with $q_2$ and $q_3$ parallel or antiparallel
remain. However the antiparallel is unlike. In fact,  as it can be
seen from Fig. \ref{fig:3}, in the antiparallel configuration we
have two $u$ quarks in the same ring reducing the phase space, and
correspondingly the gap, due to the Fermi statistics. This
observation is indeed verified by numerical calculations.
Then, one has to minimize with respect to the gap and  $\mu_e$ in
order to require electrical neutrality. The results are given in
Fig. \ref{fig:4} using the same input parameters as in Section 4 for
the gCFL case. We see that below 150 $MeV$ the LOFF phase is favored
over the normal phase with a gap arriving at almost 0.4 the CFL gap.
Of course, it is interesting to compare this result with the gCFL
result given in Fig. \ref{fig:2}. The comparison is made in Fig.
\ref{fig:5}. We see that the LOFF phase dominates over  gCFL  in the
interval between 128 $MeV$ and 150 $MeV$ where  the transition to
the normal phase is located. These results have been confirmed by an
exact calculation with respect to the gap (but always at the leading
order in the chemical potential), done in Ref.
\cite{Mannarelli:2006fy}\,. The result found by these authors show
that in the range of $M_s$ considered here the Ginzburg-Landau
approximation is rather accurate and if any it overestimates the
free energy. As a further confirmation of these results, in Ref.
\cite{Casalbuoni:2006zs} we have shown that corrections at the order
1/$\mu$ do not modify qualitatively the previous results but rather
tend to enlarge the window where LOFF dominates over gCFL.


It has also been shown in Ref. \cite{Ciminale:2006sm} that in the
phase studied in this Section the chromo-magnetic instability
disappears. Here one has to distinguish the longitudinal and
transverse masses of the gluons with respect to the direction of the
total momentum of the pair. It results that all these masses are
real.

More recently an extension of the simple ansatz of a single plane
wave for each gap, as considered in this Section, has been made in
Ref. \cite{Rajagopal:2006ig}\,. The simple ansatz of eq.
(\ref{simple_ansatz}) has been generalized in the following way \be
\langle ud\rangle\approx  \Delta\sum_a e^{2i\vec q_3^{\,a}\cdot \vec
r},~~~\langle us\rangle\approx\Delta\sum_a e^{2i\vec q_2^{\,a}\cdot
\vec r}~~~\langle ds\rangle\approx 0\ee
\begin{figure}[h]\begin{center}
\includegraphics[height=.3\textheight]{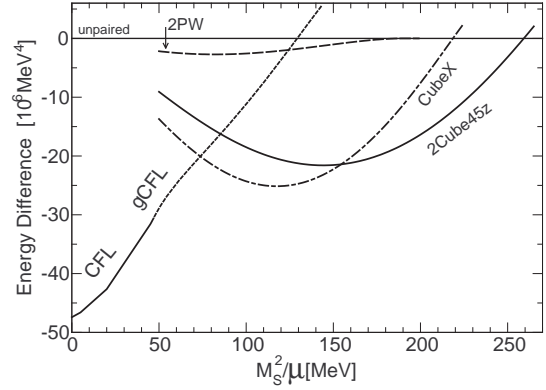}
\caption{Comparison of the free energy of the various phases already
considered in Figs. \ref{fig:2} and \ref{fig:5} (same notations as
here) with  various crystalline structures in the three flavor case.
\label{fig:6}}\end{center}
\end{figure}
with the index $a$ running from 1 up to a maximum value of 8. In
practice for any choice of the range of the index $a$ one gets a
particular crystalline structure defined by the vectors $\vec
q^{\,a}$ pointing at the vertices of the crystal.  In ref.
\cite{Rajagopal:2006ig} the study has been extended to 11 crystals.
The favored structures are the so called CubeX and 2Cube45z. The
CubeX is a cube characterized by 4 vectors $\vec q_2^{\,a}$ and 4
$\vec q_3^{\,a}$. Each set of vectors lies in a plane and the two
planes cut at 90 degrees forming a cube. In the 2Cube45z, there are
8 vectors in each set defining two cubes which are rotated one with
respect to the other of 45 degrees along the $z$ axis. The free
energies for these two crystals are compared with the case of a
single plane wave for each pairing (called in this context 2PW) in
Fig. \ref{fig:6}. We see that the CubeX and the 2Cube45z take over
the gCFL phase in almost all the relevant range of $M_s^2/\mu$.
Taking into account that this calculation has been made in the
Ginzburg-Landau approximation it looks plausible that these two
phases are the favorite ones up to the CFL phase.

\section{Conclusions}

As we have seen there have been numerous attempts in trying to
determine the fundamental state of QCD under realistic conditions
existing inside a compact stellar objects, that is to say,
neutrality in color and electric charge, $\beta$-equilibrium and a
non vanishing strange quark mass. Many competing phases have been
found. Most of them have fermionic gapless modes. However, gapless
modes in presence of a homogeneous condensate seem to lead
unavoidably to a chromo-magnetic instability and it seems necessary
to consider space dependent condensates. In this respect the LOFF
phase, where the space dependence comes about  in relation to the
non zero total momentum of the pair, seems to be a natural
candidate. This phase in the presence of three flavors has been
recently considered
\cite{casalbuoni_loff3,Mannarelli:2006fy,Rajagopal:2006ig}\,. It has
been found that there are no chromo-magnetic instabilities
\cite{Ciminale:2006sm} and that energetically it is favored almost
up to the CFL phase. However, considering the approximations
involved in these calculations, before to draw sounded conclusions
one should  attend for more careful investigations.

%\newpage
%\bibliographystyle{ws-procs9x6}
%\bibliography{ws-pro-sample}

\end{document}